\documentclass[twocolumn,trackchanges]{aastex701}
\usepackage{amsmath}
\usepackage{caption}

\begin{document}

\title{Implications of a Cosmogenic Origin of KM3-230213A for Ultra-High-Energy Protons}

\author[orcid=0009-0002-7320-7638]{Abdulrahman Alhebsi}
\affiliation{Khalifa University of Science and Technology, Department of Physics, PO Box 127788, Abu Dhabi, UAE}
\email[show]{100064470@ku.ac.ae}  

\author[orcid=0000-0003-2827-3361]{Arjen van Vliet} 
\affiliation{Khalifa University of Science and Technology, Department of Physics, PO Box 127788, Abu Dhabi, UAE}
\email{arjen.vliet@ku.ac.ae}  

\author[orcid=0000-0002-4322-6400]{Domenik Ehlert}
\affiliation{Norwegian University of Science and Technology (NTNU), 7491 Trondheim, Norway}
\email{domenik.ehlert@ntnu.no}

\author[orcid=0000-0002-7066-3614]{Satyendra Thoudam} 
\affiliation{Khalifa University of Science and Technology, Department of Physics, PO Box 127788, Abu Dhabi, UAE}
\email{satyendra.thoudam@ku.ac.ae}

\correspondingauthor{Abdulrahman Alhebsi}

\begin{abstract}
A significant neutrino event with an estimated energy between 72$\,\mathrm{PeV}$ and 2.6$\,\mathrm{EeV}$ was recently observed by the KM3NeT experiment (KM3-230213A). When interpreted as cosmogenic in origin, this event can provide constraints on several phenomenological parameters of ultra-high-energy (UHE) proton sources. In this study, we present the best fit to the spectrum and composition of ultra-high-energy cosmic rays (UHECRs) that is consistent with multi-messenger constraints, including the detection of a single neutrino event by the KM3NeT detector at the energy range of KM3-230213A. From the best fit, we obtain the 68\% CL constraints on the parameters of a two-population model of UHECRs, comprising a mixed-composition population and a subdominant UHE proton population. Our results indicate that the detection of a single neutrino event at the energy range of KM3-230213A solely with the KM3NeT exposure requires strongly evolving UHE proton sources, consistent with high-luminosity active galactic nuclei. On the other hand, including the null observations from the Pierre Auger and IceCube observatories disfavors such strong evolution. In both cases, the observed proton fraction of UHECRs is primarily constrained by the composition data to be ${\sim\,} 20\%$ at 20$\,\mathrm{EeV}$.
\end{abstract}


\section{Introduction}
Ultra-high-energy (UHE) cosmic rays (CRs), with energies ${\gtrsim\,} 10^{18}\,\mathrm{eV}$, are deflected by Galactic and extragalactic magnetic fields, making it challenging to trace them back to their sources.
Compounding this challenge are their limited statistics due to their extremely low flux at the highest energies and their unclear composition due to uncertainties in interpreting measurements of the depth of air shower maximum, $X_{\text{max}}$. These challenges motivate combining UHECR data with multi-messenger signals to understand the sources of UHECRs.
\par \smallskip

UHECRs can be studied by examining the cosmogenic neutrino and photon signals they generate through interactions with low-energy photons of the cosmic microwave background (CMB) and extragalactic background light (EBL) during intergalactic propagation. In particular, the observed flux of cosmogenic neutrinos at ${\sim\,} 1\,\mathrm{EeV}$ is largely determined by two properties of UHECRs: (1) the proton fraction of cosmic rays observed at ${\sim\,} 20\,\mathrm{EeV}$ and (2) the cosmological evolution of the sources of these UHE protons. This follows from the fact that, at energies ${\gtrsim\,} 20\,\mathrm{EeV}$, protons primarily lose energy through photo-pion production, with the decay of the resulting charged pions yielding neutrinos that may contribute significantly to the cosmogenic neutrino flux. Furthermore, the cosmological evolution of the emissivity of UHE proton sources determines the number of interactions that occur at high redshift. If the emissivity of proton sources evolves strongly with redshift, more interactions occur at earlier cosmic times, which directly increases the cosmogenic neutrino flux observed on Earth. Hence, an exciting prospect is to combine the observed cosmogenic neutrino flux with UHECR data to place constraints on the observed fraction of UHE protons beyond the ankle ($10^{18.7}\,\mathrm{eV}$) and the cosmological evolution of their sources. This possibility was noted by \citet{takami2009cosmogenic}, \citet{ahlers2009neutrino} and \citet{globus2017probing}, and later investigated in detail by \citet{arjen_fp_2019}, \citet{muzio2023prospects}, and \citet{ehlert2024constraints}.
\par 
\smallskip

The cosmological evolution of UHECR sources remains largely undetermined. Nevertheless, the observed fluxes of secondary CRs, as well as cosmogenic gamma rays and neutrinos, set an upper limit on the number of interactions that UHECRs undergo in intergalactic space, and thus on how strongly the UHECR source emissivity can increase with redshift. For a mixed mass composition of UHECRs, for example, a source emissivity evolving with redshift as $\propto (1+z)^5$ or steeper, produces a flux of low-energy secondary CRs that exceeds the observed UHECR spectrum below the ankle \citep{batista2019cosmogenic, halim2023constraining}. Comparable constraints have also been obtained from diffuse gamma-ray and neutrino upper limits for different composition scenarios of UHECRs, such as mixed-composition scenarios \citep{globus2017probing, halim2023constraining}, scenarios with a subdominant proton-producing population \citep{muzio2023prospects, ehlert2024constraints}, and the proton dip scenario \citep{heinze2016cosmogenic, berezinsky2016cascade}. Because the cosmological evolution varies widely between known source classes, characterizing it for UHECR sources constrains the pool of candidate sources.
\par \smallskip

The proton fraction of observed UHECRs also remains uncertain, though to a lesser extent than the cosmological evolution of their sources. Information on the mass composition of UHECRs is primarily inferred from extensive air-shower measurements, in particular the distribution of the depth of shower maximum, $X_{\text{max}}$, in each energy bin. Because interpreting this observable relies on hadronic interaction models, composition inferences carry systematic uncertainties associated with the choice of model \citep{abdul2024testing}. The Pierre Auger collaboration has interpreted their observed $X_{\text{max}}$ distributions using different hadronic interaction models \citep{auger_xmax_2014, auger_xmax_icrc23} and found that all models agree on a very light composition at ${\sim\,} 3\,\mathrm{EeV}$, which becomes heavier with increasing energy. Although the absolute proton fraction differs between models, they predict a similar trend in the proton fraction with energy, which is characterized by a decrease from ${\sim\,} 40{-}60\%$ at ${\sim\,} 3\,\mathrm{EeV}$ to ${\sim\,} 0{-}20\%$ at energies ${\gtrsim\,} 10\,\mathrm{EeV}$, within systematic uncertainties. Complementary to air-shower measurements, knowledge of the UHECR composition can be inferred from multi-messenger signals. For a fixed source evolution, cosmogenic neutrino measurements set an upper limit on the number of photo-pion interactions that UHE protons undergo in intergalactic space, and thus on the fraction of protons in the observed UHECR flux. Upper limits on the proton fraction at $10^{19.55}\,\mathrm{eV}$ were obtained for a range of evolution strengths of the source emissivity using upper limits on the UHE (${\gtrsim\,} 10^{16}\,\mathrm{eV}$) diffuse neutrino flux set by the Pierre Auger \citep{auger_nu_null} and IceCube \citep{IC_UL2025} observatories.
\par \smallskip

In this work, we aim to obtain new constraints on the observed fraction of UHE protons and the cosmological evolution of their sources in light of the UHE neutrino event, KM3-230213A \citep{km3_nature}, under the assumption that the event originated from a steady diffuse flux of cosmogenic neutrinos. For this purpose, we use a two-population model of UHECRs adapted from \citet{ehlert2024constraints}, comprising a mixed-composition population and a subdominant UHE proton population. Using this model, we perform a joint fit of the spectrum \citep{auger_spec2021} and composition observables \citep{auger_fd_icrc23} of UHECRs reported by the Pierre Auger Observatory (Auger) while satisfying the detection of a single neutrino event by the Cubic Kilometer Neutrino Telescope (KM3NeT) at the energy range of KM3-230213A, neutrino constraints from IceCube and Auger, and gamma-ray constraints from Auger and the Large Area Telescope on board the \textit{Fermi} gamma-ray space telescope (\textit{Fermi}-LAT). From the best-fit solution, we obtain the 68\% CL constraints on the model parameters.
\par \smallskip

This paper is structured as follows. In Section~\ref{sec: model}, we describe the model. In Section~\ref{sec: MM datasets}, we present the multi-messenger datasets used in this analysis and discuss the UHE neutrino event, KM3-230213A. In Section~\ref{sec: method}, we present the method of this analysis and the fit procedure. In Section~\ref{sec: results}, we present the main results and provide the best-fit parameters. In Section~\ref{sec: discussion}, we discuss the viability of the assumptions adopted in this work, and discuss recent studies closely related to our analysis. Lastly, in Section~\ref{sec: conclusion}, we present the major conclusion of this work. 
\par \smallskip

\section{Model}\label{sec: model}

\subsection{Cosmic-ray Injection Spectra}

Our model comprises two independent populations of UHECRs whose sources are distributed homogeneously in space. The first population is mixed in composition with five nuclear species: $^1$H, $^4$He, $^{14}$N, $^{28}$Si, and $^{56}$Fe. The second population is composed purely of protons. For each
population, the CR spectrum observed at Earth is dependent on the spectrum of CRs injected into space, which we parameterize as:
{\small
\begin{equation}\label{J_esc_eq}
    J_{\text{inj}}(E) = J_0 \sum_A f_A \left(\frac{E}{E_0}\right)^{-\gamma} \exp\left(-\frac{E}{Z_A R_{\text{max}}}\right)
\end{equation}
}
Here, $\gamma$ and $R_{\text{max}}$ are the spectral index and the maximum rigidity of the injected spectrum, respectively, $J_0$ is a free normalization parameter, and $E_0$ is an arbitrary normalization energy. Each nuclear species is represented with its atomic mass $A$, where $Z_A$ denotes its atomic number, and $f_A$ denotes its compositional fraction at a fixed energy with respect to the total injected spectrum. For the pure-proton population $\sum_A f_A = f_p = 1$. It is important to note that the parameters for the two populations are independent.
\par \smallskip

The two-population model we use is adapted from the model presented in \cite{ehlert2024constraints}. Similar to previous studies \citep{muzio2019progress, muzio2023prospects, das2021modeling}, this work showed that including a subdominant proton component at the highest energies improves fits to the UHECR data, given the assumptions adopted in these studies. This scenario is consistent with interpretations of the $X_{\text{max}}$ distributions using different hadronic interaction models \citep{auger_xmax_2014, auger_xmax_icrc23}, all of which predict a subtle increase in the proton fraction around ${\sim\,} 20\,\mathrm{EeV}$, indicating a possible recovery in the flux of UHE protons. In fact, the data allow for a proton fraction as high as 20\% at this energy. These results motivate adopting a model that allows the proton fraction to vary independently of the mixed-composition population. Furthermore, the presence of a subdominant proton population allows the model to reproduce a wide range of cosmogenic neutrino fluxes in light of new observations, while remaining consistent with the UHECR data. On the other hand, typical single-population models exhibit an increasingly heavy composition towards higher energies as a consequence of the Peter's cycle assumption, where a common maximum rigidity is assumed for all nuclei \citep{peters1961primary}, resulting in a relatively low cosmogenic neutrino signal \citep[e.g., ][]{batista2019cosmogenic, heinze2019new}.
\par \smallskip

Compared to the study presented in \cite{ehlert2024constraints}, which focused primarily on energies above the ankle, we extend this analysis to lower energies by incorporating an independent CR component below the ankle. This flux could be provided either by a population of high-energy Galactic sources or by a population of extragalactic sources independent of the mixed-composition population providing the flux above the ankle. For the subsequent analysis, we adopt a Galactic CR flux model from \citet{thoudam2016cosmic}, in particular, the CR component resulting from supernova explosions in the environment of Wolf-Rayet stars, which dominates at energies between the second knee ($10^{17}\,\mathrm{eV}$) and the ankle. In our analysis, we consider the normalization of the Galactic component as a free parameter, while keeping the spectral index, the energy cutoff, and the composition fixed at the values reported in \citet{thoudam2016cosmic}.
\par \smallskip

\subsection{Cosmological Evolution of Sources}\label{sec: source evolution}

The UHECR spectrum observed at Earth also depends on the cosmological evolution of UHECR sources, which we parameterize in our model as:
{\small
\begin{equation}\label{eq: SE}
    S(z) =
    \begin{cases}
        (1 + z)^m & \text{for } m \leq 0 \text{ and }  z \leq z_{\text{max}}\\
        (1 + z)^m & {\text{for } m > 0 \text{ and } z \leq z_0} \\
        (1 + z_0)^m & \text{for } m > 0 \text{ and } z_0 < z \leq z_{\text{max}} \\
        0 & \text{for } z>z_{\text{max}}.
    \end{cases}
\end{equation}
}
This function describes how the source emissivity changes with redshift, where $m$ determines the strength and direction (positive or negative) of the evolution. This parameterization, proposed in \citet{arjen_fp_2019}, was chosen because typical source candidates show a growth in emissivity up to a redshift $1 \lesssim z \lesssim 1.7$, reach a plateau or increase very slowly up to a redshift $2.7 \lesssim z \lesssim 4$, and then decline beyond that. This behavior is observed in gamma-ray bursts \citep[GRBs;][]{GRB_m}, the star-formation rate \citep[SFR;][]{SFR_m_value}, and AGNs of different luminosity classes \citep{AGN_m_value, seyfert_m}. In Equation~\ref{eq: SE}, $z_0$ denotes the redshift beyond which the source emissivity remains constant, and $z_{\text{max}}$ denotes the redshift to which the source population extends. In this analysis, we fix $z_0$ and $z_{\text{max}}$ to 1.5 and 4, respectively. Since sources differ in the strength of their cosmological evolution, the source-evolution parameter, $m$, allows us to describe a wide range of sources. For example, specific classes of tidal disruption events can have a negative evolution close to $m {\,\sim} -3$ \citep{TDE_m}, whereas high-luminosity AGNs, such as flat-spectrum radio quasars (FSRQs) and high-luminosity BL Lacertae objects (BL Lacs), have strongly positive evolutions of $m {\,\sim\,} 5{-}7$ \citep{FSRQ_m, HL_BLlacs_m}.
\par \smallskip

\section{Multi-messenger Datasets}\label{sec: MM datasets}
The multi-messenger datasets and upper limits used in this study are the following: (1) IceCube's high-energy event samples, namely High-Energy Starting Events \citep[HESE;][]{HESE2021}, Northern-Sky Tracks \citep[NST;][]{NST2022}, and the Glashow resonance event \citep{Glashow2021}; (2) IceCube's Extremely-High-Energy (IC-EHE) sample \citep[null observations in 12.6 years of livetime,][]{IC_UL2025}; (3) Auger's null observations of UHE neutrinos \citep[livetime of 18 years;][]{auger_fp_constr_aeff}; (4) 95\% CL upper limits on the diffuse flux of UHE (${\gtrsim\,} 10^{18}\,\mathrm{EeV}$) gamma rays set by Auger \citep{auger_UHE_gamma_UL, auger_UHE_gamma_UL_2024}; (5) the isotropic diffuse gamma-ray background (IGRB) measured by the \textit{Fermi}-LAT \citep{Fermi-LAT_IGRB2015}; and (6) KM3-230213A \citep{km3_nature}.
\par \smallskip

Until recently, no neutrino candidates with energies above ${\sim\,} 10^{16}\,\mathrm{eV}$ were detected. This changed on February 13, 2023, when the KM3NeT Astroparticle Research with Cosmics in the Abyss (ARCA) neutrino telescope, currently under construction in the Mediterranean Sea, detected the highest-energy neutrino candidate--- KM3-230213A. This nearly horizontal-going neutrino event was reconstructed with an energy of approximately 0.220$\,\mathrm{EeV}$, with a 90\% CL interval of $0.072 {-}2.6\,\mathrm{EeV}$. Given the event’s energy and direction, along with the steep decline of the atmospheric neutrino spectrum at PeV energies \citep{ostapchenko2023prompt}, an astrophysical origin of the event is strongly favored.
\par \smallskip

While the exact origin of KM3-230213A remains unidentified, different origin scenarios have been investigated, which include a population of steady neutrino emitters, such as blazars \citep[e.g.,][]{km3net2025blazars}, a single flaring blazar \citep[e.g.,][]{km3_direction_companion, km3_transient_jet, dzhatdoev2025blazar}, other transient phenomena \citep[e.g.,][]{km3_muon_burst, mukhopadhyay2026high, neronov2026km3}, cosmogenic production \citep[e.g.,][]{km3_cosmogenic_companion, boxi2026cosmogenic, zhang2025cosmogenic,muzio2025emergence, kuznetsov2026ultra, km3_alessandro}, a galactic origin \citep[e.g.,][]{km3_galactic}, or heavy dark matter decay \citep[e.g.,][]{km3_darkmatter_borah, km3_darkmatter_kohri, murase2025superheavy}. The hypothesis that KM3-230213A originates from a steady diffuse astrophysical component not yet identified by existing experiments cannot be ruled out, given that the probability of observing one event by KM3NeT and none by the Pierre Auger and IceCube observatories is about 0.4\% when considering the combined exposure of all three experiments to UHE neutrinos \citep{IC_UL2025}. Moreover, directional associations of KM3-230213A to blazar candidates do not rule out a cosmogenic origin scenario. This is because high-rigidity CRs escaping their source environments close to our line of sight can interact with background photons before getting significantly deflected by intergalactic magnetic fields, creating `line-of-sight' cosmogenic neutrinos \citep{essey2010secondary, essey2011role}, as was hypothesized for the case of this event \citep{das2025cosmic} and GRB 221009A \citep[see e.g., ][]{batista2022grb, das2023ultrahigh}.
\par \smallskip

Because we are concerned with the cosmogenic component of astrophysical neutrinos, the multi-messenger datasets that we aim to explain with our model are KM3-230213A and neutrino upper limits reported by IceCube and Auger above ${\sim\,} 5 \,\mathrm{PeV}$, which we shall refer to as the UHE neutrino datasets. The remaining datasets, which include the neutrino flux below ${\sim\,} 5 \,\mathrm{PeV}$ from IceCube's high-energy event samples, the IGRB flux, and Auger's upper limits on the UHE gamma-ray flux, serve only as limits not to be exceeded (see Section~\ref{sec: method}). In particular, we constrain the model-predicted GeV$-$TeV gamma-ray flux to contribute no more than 40\% of the IGRB, since unresolved gamma-ray sources are predicted to account for at least the remaining 60\% \citep{IGRB_68contrib, IGRB_86contrib}.
\par \smallskip

\begin{figure*}[t]
    \centering
    \begin{minipage}{0.5\textwidth}
    \includegraphics[width=\linewidth]{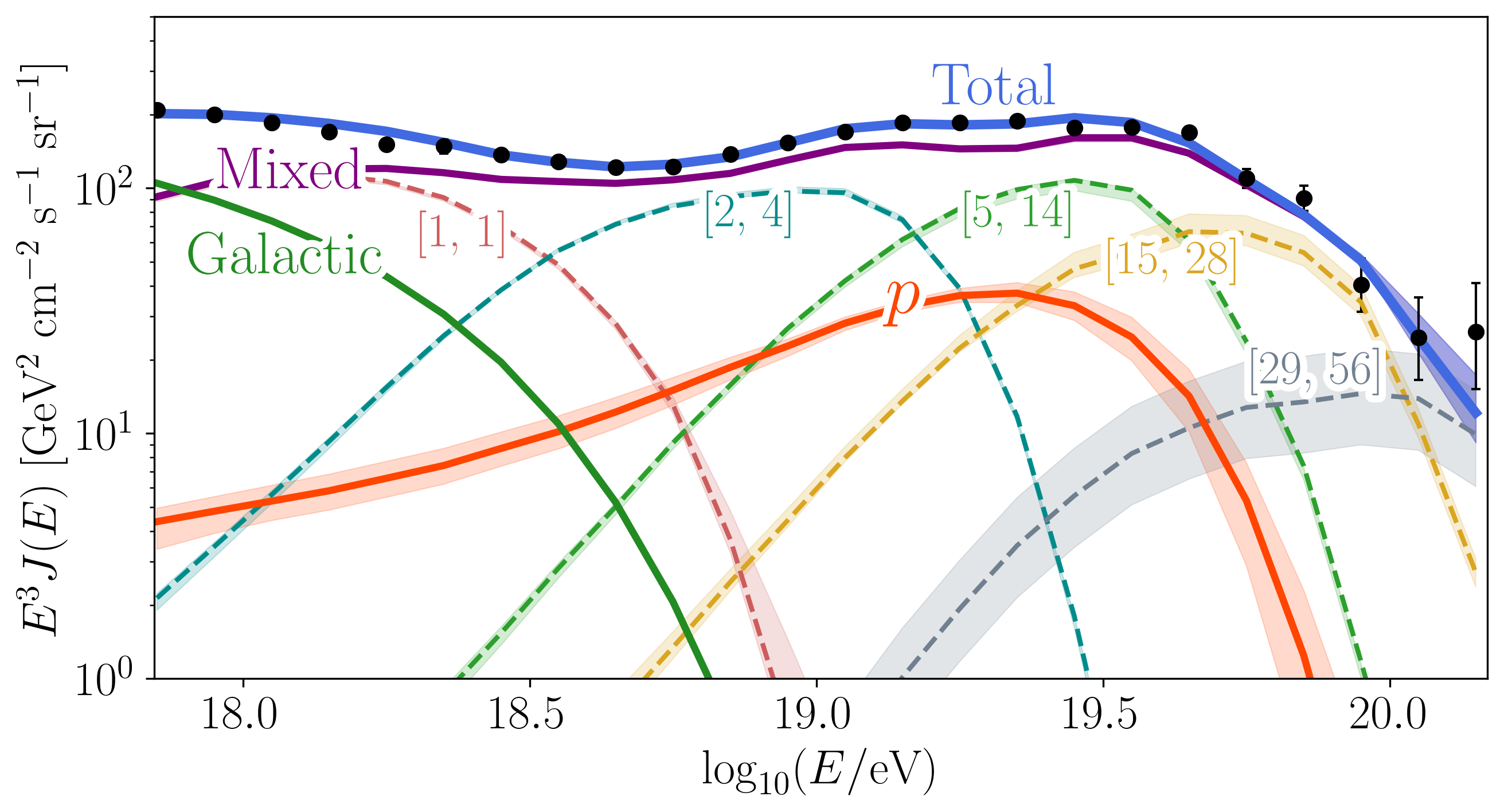}
    \includegraphics[width=\linewidth]{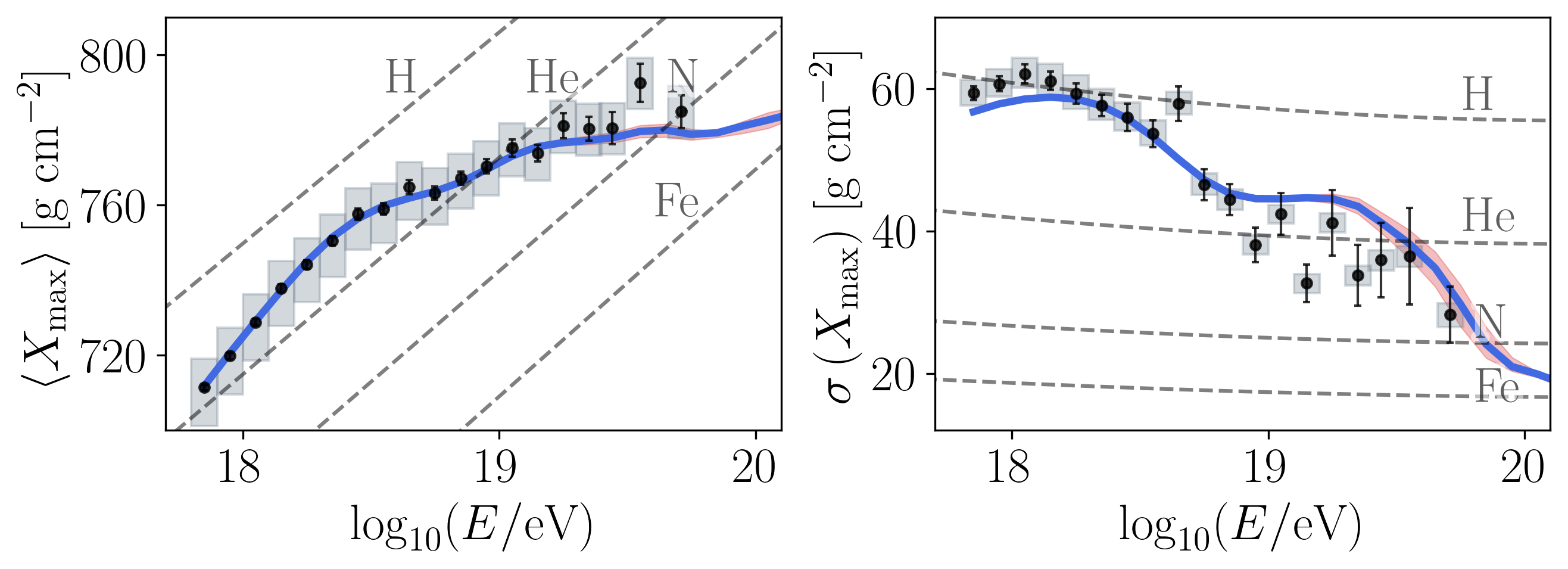}
    \end{minipage}
    \hspace{-0.3cm}
    \begin{minipage}{0.5\textwidth}
    \includegraphics[width=\linewidth]{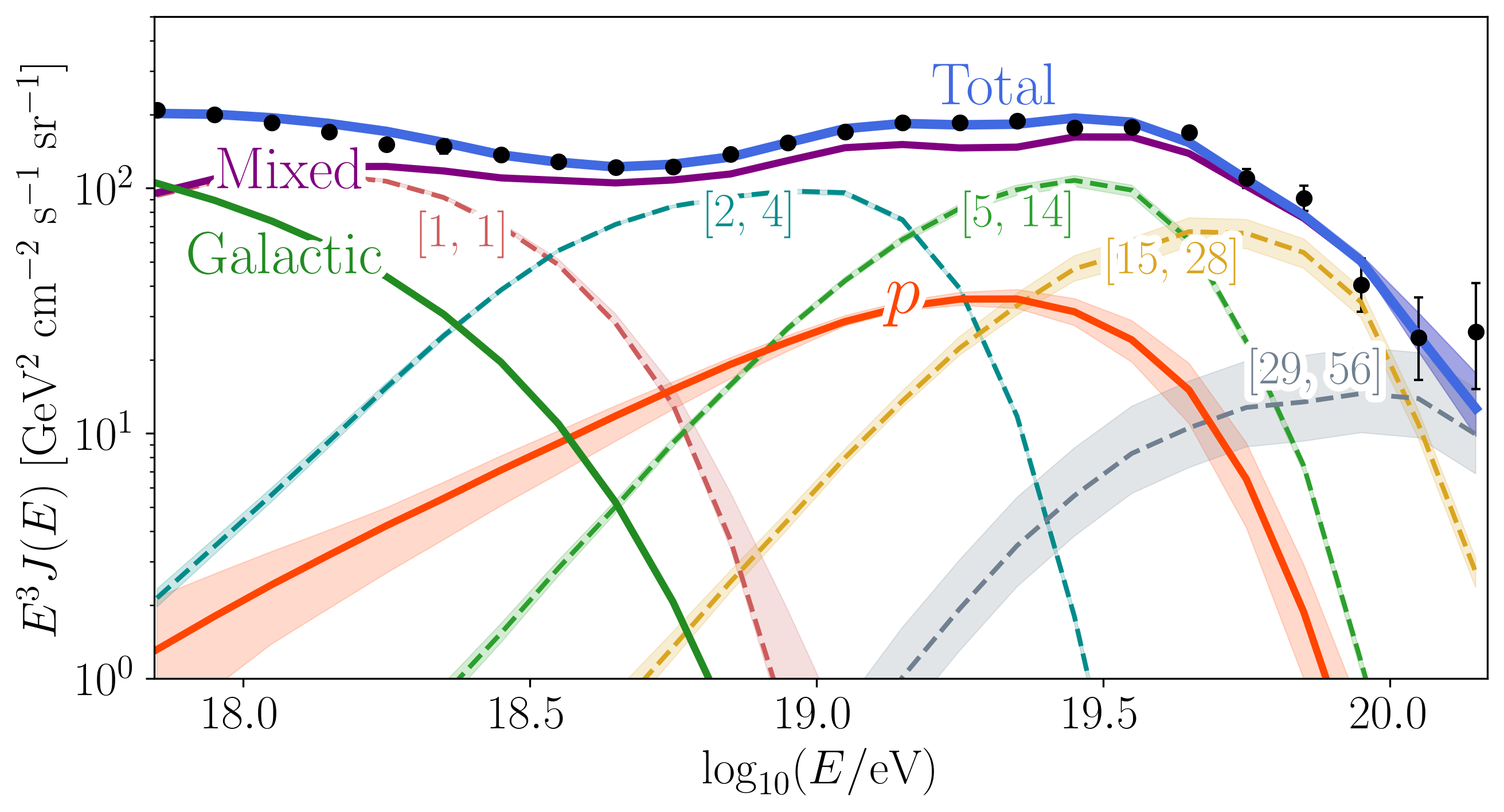}
    \includegraphics[width=\linewidth]{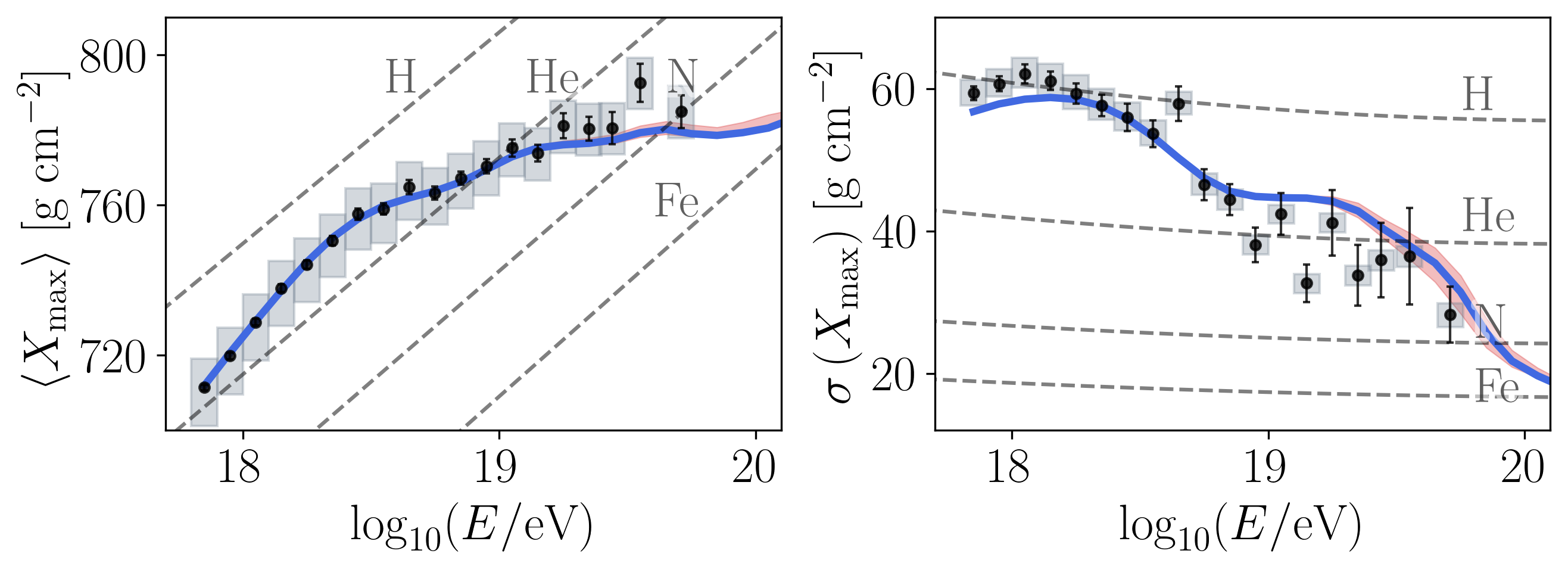}
    \end{minipage}
    \captionsetup{width=0.9\linewidth}
    \caption{The CR spectrum (top) and composition observables (bottom) of the KM3NeT-only (left) and joint (right) fits shown alongside data by Auger \citep{auger_spec2021, auger_fd_icrc23}. The contribution of different mass groups is shown with $[A_{\text{min}},  A_{\text{max}}]$, and colored bands indicate the 68\% CL range. The Galactic component is added to our model from \cite{thoudam2016cosmic}, albeit with a free
normalization. In the lower panels, the \textsc{epos-lhc} hadronic interaction model is used to interpret different nuclei (grey dashed lines) and the model-predicted composition (solid blue line) into $\langle X_{\text{max}}\rangle$ and $\sigma(X_{\text{max}})$. }
    \label{fig: CR_fit}
\end{figure*}

\section{Method}\label{sec: method}
We use the simulation framework \textsc{CRPropa 3.2} \citep{Alves_Batista_2022} to produce one-dimensional Monte Carlo events of propagated UHECRs and their secondary cosmogenic neutrinos and gamma rays. The propagation of secondary photons and electrons subsequently produces an electromagnetic cascade that results in a diffuse flux of GeV$-$TeV gamma rays. Because simulating such cascades is computationally expensive, we resort to the thinning feature in \textsc{CRPropa}, where a smaller number of representative particles are added to the list of secondaries with weights assigned to them. We use the EBL model from \citet{gilmore2012semi} and the default photodisintegration cross sections implemented in \textsc{CRPropa} \citep[see][for details]{kampert2013crpropa, batista2016crpropa}, which are partially obtained from the TALYS 1.8 model \citep{koning2012modern}. We discuss the uncertainties associated with the choice of models in Section \ref{sec: discussion}. The composition observables we fit are the mean $\langle X_{\text{max}} \rangle$ and the standard deviation $\sigma(X_{\text{max}})$ of the $X_{\text{max}}$ distribution in each energy bin \citep{auger_fd_icrc23}. To compare our model with these observables, we convert the average atomic mass in each energy bin into these quantities using the parameterization of \cite{abreu2013interpretation} for the \textsc{epos-lhc}\footnote{Parameter values were obtained through private communication with S. Petrera.} hadronic interaction model.
\par \smallskip

We fit the data by minimizing the deviance, $D= -2\ln(L/L_{\text{sat}})$, which is equivalent to maximizing the model's likelihood, $L$, where $L_{\text{sat}}$ is the likelihood of a saturated model that perfectly describes the data. For the UHECR spectrum and composition, we assume Gaussian likelihood distributions:
\begin{equation}\label{deviance_cr}
    D_{\text{CR}} = \sum_F\sum_i \left[\frac{F_{\text{Auger}}(E_i) - F_{\text{model}}(E_i; \mathbf{P})}{\sigma_i} \right]^2 ,
\end{equation}
computed for a set of parameter values, $\mathbf{P}$, for three Auger observables $F \in \{J_{\text{CR}}, \,\langle X_{\text{max}} \rangle, \allowbreak  
 \sigma(X_{\text{max}})\}$ at each $i$-th energy bin in $E\geq 10^{17.85}\,\mathrm{eV}$. For the UHE neutrino datasets, we assume Poissonian likelihood distributions:
\begin{equation}\label{deviance_nu}
    D_{\text{UHE }\nu} = 2\sum_d \sum_i \mu^i_{d} - n^{i}_{d} + n^{i}_{d}\ln(n^{i}_{d}/\mu^{i}_{d})
\end{equation}
where $\mu^i_d$ and $n^i_d$ are the expected
and observed number of events, respectively, at the $i$-th energy bin of dataset $d \in \{$Auger, KM3NeT, IC-EHE, HESE($> 5\,\mathrm{PeV}$), NST($> 5\,\mathrm{PeV}$)$\}$. For energy bins with zero observed events, i.e., outside the 90\% CL interval of KM3-230213A, Equation \ref{deviance_nu} simplifies to $2\sum_d \mu_d$, where $\mu_d$ is given by:
\begin{equation}
    \mu_d = \frac{1}{3} \int^{E_{\text{max}, d}}_{E_{\text{min}, d}} \mathcal{E}^{\text{all-flavor}}_d (E)\times \Phi^{\text{all-flavor}}_{\nu+\bar{\nu}}(E) \; dE
\end{equation}
where $\phi^{\text{all-flavor}}_{\nu+\bar{\nu}}$ is the model-predicted all-flavor neutrino flux and  $\mathcal{E}^{\text{all-flavor}}_d$ is the sky-averaged, all-flavor detector exposure corresponding to dataset $d$. Both quantities are averaged between neutrinos and anti-neutrinos. The integration is carried over the energy range of the dataset $[E_{\text{max}, d}, {E_{\text{min}, d}}]$, and the result is multiplied by $1/3$ to return the expected number of events from a single neutrino flavor. The detector exposures of Auger, IceCube, and KM3NeT/ARCA used to calculate Equation~\ref{deviance_nu} were obtained from \citet{auger_fp_constr_aeff}, \citet{IC_UL2025}, and \citet{km3_landscape_companion}, respectively.
\par \smallskip

 \begin{figure*}[t!]
    \centering
    \includegraphics[width=\linewidth]{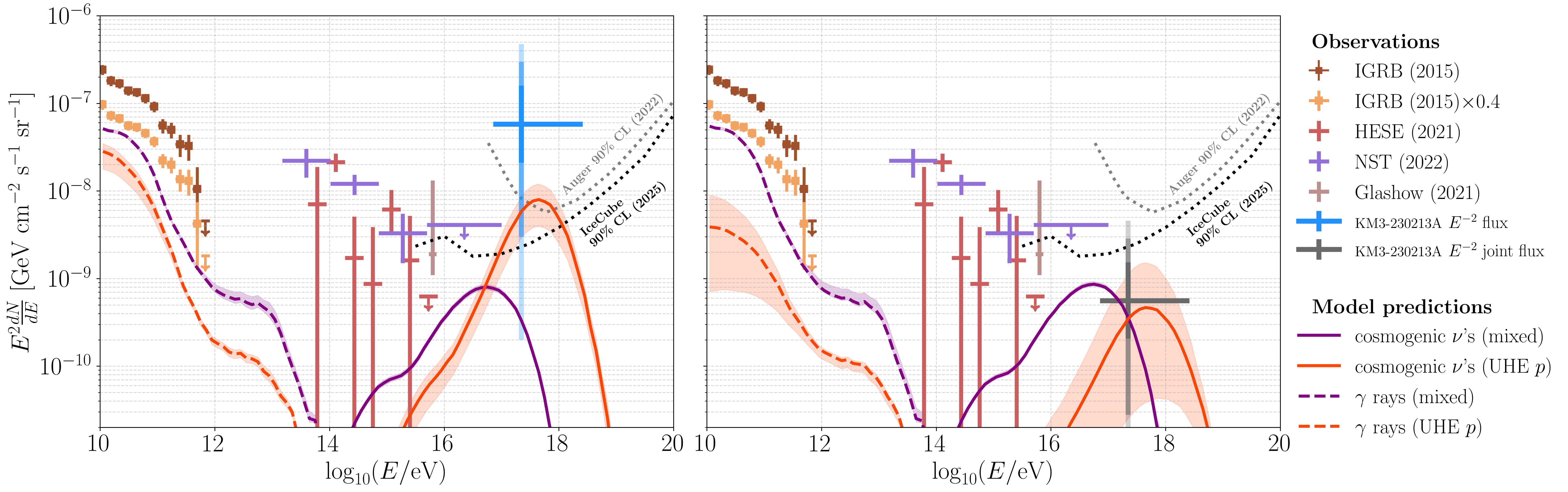}
    \captionsetup{width=0.9\linewidth}
    \caption{Cosmogenic neutrinos (solid lines) and diffuse electromagnetic-cascade gamma rays (dashed lines) of the KM3NeT-only (left) and joint (right) fits. All neutrino fluxes are per-flavor, assuming flavor equipartition post-propagation. Contributions from the mixed and pure-proton populations are shown in purple and orange, respectively, with bands indicating the 68\% CL range. We show the IGRB measured by \textit{Fermi}-LAT, IceCube's high-energy event samples (High-Energy Starting Events (HESE), Northern-Sky Tracks (NST), and the Glashow resonance event), as well as the 90\% CL neutrino upper limits reported by IceCube and Auger. We also show the $E^{-2}$ flux associated with KM3-230213A (see main text), for which the horizontal error bar represents the 90\% CL interval on the estimated neutrino energy, and the vertical error bars represent the 68\%, 95\%, and 97.5\% Feldman-Cousins confidence intervals \citep{feldman1998unified}.}
    \label{fig: MM}
\end{figure*}

For the neutrino flux below ${\sim\,} 5 \,\mathrm{PeV}$ from IceCube's high-energy event samples, the IGRB flux, and Auger's upper limits on the UHE gamma-ray flux, we use a one-sided Gaussian likelihood deviance that contributes only for energy bins where the predicted flux exceeds the average value of the observed flux, thereby acting as a penalty term which we denote by $D_{\text{penalty}}$. The total deviance for the minimization then becomes $D = D_{\text{CR}} + D_{\text{UHE }\nu} +D_{\text{penalty}}$.
\par \smallskip

\section{Results}\label{sec: results}
The best-fit solution is determined for two scenarios. In the first scenario, neutrino upper limits from Auger and IceCube are disregarded, and Equation~\ref{deviance_nu} is evaluated using only the KM3NeT dataset. We refer to this as the KM3NeT-only fit. In the second scenario, Equation~\ref{deviance_nu} is evaluated for all three detectors, and we refer to this as the joint fit.
\par \smallskip

Figure~\ref{fig: CR_fit} shows the spectrum (upper panels) and composition observables (lower panels) of UHECRs arriving at Earth as reported by Auger \citep{auger_spec2021, auger_fd_icrc23}, together with model predictions of the KM3NeT-only (left panels) and joint (right panels) fits. We have divided the mixed-composition population in our model into five components corresponding to the five injected species of $^1$H, $^4$He, $^{14}$N, $^{28}$Si, and $^{56}$Fe. The other two components shown are the Galactic CRs and the subdominant UHE proton population. Both fits predict a relatively hard spectral index of the injected spectra for both the mixed-composition and the pure-proton populations. In both populations, the flux suppression is dominated by the maximum rigidity cutoff at the sources.
\par \smallskip

\begin{figure}[t]
    \centering
    \includegraphics[width=\linewidth]{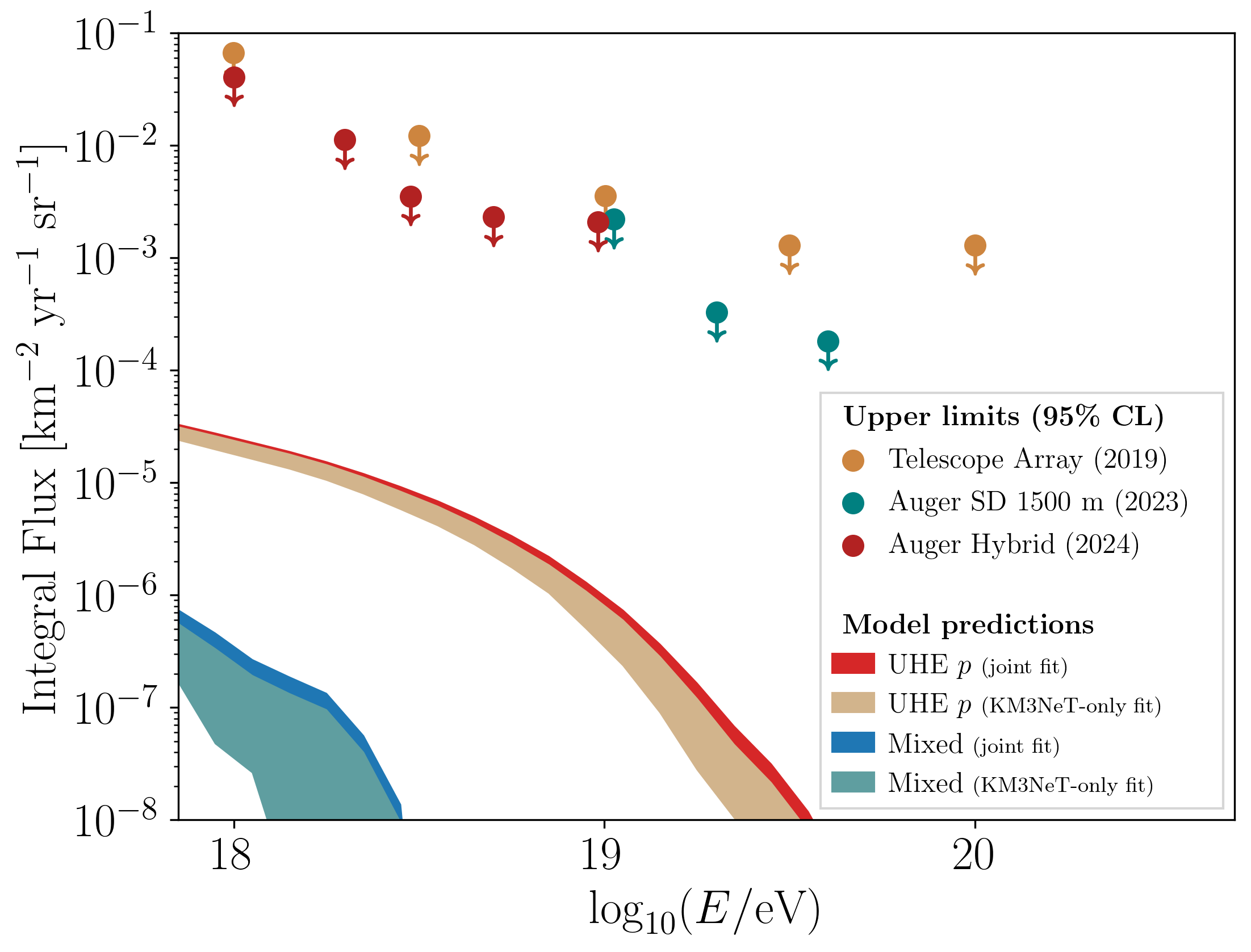}
    \caption{The 68\% CL range in the model-predicted integral flux of UHE gamma rays and current 95\% CL upper limits posed by Auger \citep{auger_UHE_gamma_UL, auger_UHE_gamma_UL_2024} and the Telescope Array \citep{TA_UHE_gamma_UL}. }
    \label{fig: UHE_ga}
\end{figure}

Both fits exhibit a similar evolution of the proton fraction in the observed spectrum, peaking to ${\sim\,} 65\%$ at ${\sim\,} 2\,\mathrm{EeV}$, predominantly due to secondary protons produced by photodisintegration of heavier nuclei in the mixed-composition population. Above that energy, the contribution from the subdominant proton population increases steadily until ${\sim\,} 20\,\mathrm{EeV}$, where the proton fraction reaches a second peak of ${\sim\,} 20\%$. Beyond this point, it declines until becoming negligible at energies ${\gtrsim\,} 100\,\mathrm{EeV}$. For each fit, we report the observed proton fraction only at 20$\,\mathrm{EeV}$, which we denote by $f_p^{\text{obs}}(\text{20$\,\mathrm{EeV}$})$. The elemental fractions of the mixed-composition population before propagation are reported at a fixed rigidity, $f^R_A$, such that the ratio of the flux of element $A$ to the total flux is reported at the same rigidity for all elements\footnote{A transformation between $f^R_A$ and the fraction at a fixed energy defined in Equation~\ref{J_esc_eq} can be achieved via $f^R_A = f_A \cdot Z(A)^{1-\gamma}$, plus subsequent renormalization.}
\par \smallskip

Figure~\ref{fig: MM} shows the corresponding cosmogenic gamma-ray and neutrino spectra predicted by the KM3NeT-only (left panel) and joint (right panel) fits, plotted alongside the IGRB data by \textit{Fermi-}LAT \citep{Fermi-LAT_IGRB2015}, IceCube's high-energy event samples \citep{HESE2021, Glashow2021, NST2022}, and the UHE neutrino upper limits by Auger \citep{auger_UL2022} and IceCube \citep{IC_UL2025}. Alongside the KM3NeT-only fit, we plot the single-flavor $E^{-2}$ flux for KM3-230213A reported in \citet{km3_nature}, which is based solely on the KM3NeT/ARCA exposure. In the joint-fit panel, we plot the corresponding $E^{-2}$ flux derived from the combined exposures of Auger, IceCube, and KM3NeT/ARCA obtained from \citet{auger_fp_constr_aeff}, \citet{IC_UL2025}, and \citet{km3_landscape_companion}, respectively, which we calculate\footnote{The method for this calculation is detailed in \citep{km3_landscape_companion}, where an estimate of $E^2\Phi^{\text{1f}}_{\nu+\bar{\nu}} = 7.5^{+13.1}_{-4.7} \times 10^{-10}\,\mathrm{GeV\,cm^{-2}\,s^{-1}\,sr^{-1}}$ was obtained using the 9-year IceCube exposure \citep{IC_UL_18}, as opposed to the 12.6-year exposure used for our estimate.} to be $E^2\Phi^{\text{1f}}_{\nu+\bar{\nu}} = 5.6^{+9.8}_{-3.5} \times 10^{-10}\,\mathrm{GeV\,cm^{-2}\,s^{-1}\,sr^{-1}}$ with the 68\% Feldman-Cousins confidence interval \citep{feldman1998unified}.
\par \smallskip

\begin{table}[t]
\centering
\caption{The 68\% CL intervals around the best-fit parameters. The number of degrees of freedom are $N_{\text{dof, CR}}=49$ and $N_{\text{dof, UHE } \nu}=2$ from the two Poissonian distributions inside and outside the energy range of KM3-230213A. Confidence intervals that touch the boundaries of the parameter space are indicated by an asterisk.}
\setlength{\tabcolsep}{0.5pt}
\begin{tabular}{lcc}
\hline
\hline
\textbf{Best fit} & KM3NeT-only & Joint \\
\hline
\textbf{Mixed-composition population} & & \\
\hline
$\gamma$ & $-2.77^{+0.11}_{-0.11}$ & $-2.78^{+0.13}_{-0.13}$  \\
$R_{\max}$ [EV] & $1.23^{+0.03}_{-0.03}$ & $1.24^{+0.03}_{-0.04}$ \\
$m$ & $3.9^{+0.1}_{-0.1}$ & $4.0^{+0.1}_{-0.1}$ \\
\hline
$f^R_{\text{H}}$ [\%] & $0.2^{+6.8}_{-0.2}$ & $ 0.3^{+9.5}_{-0.3}$ \\
$f^R_{\text{He}}$ [\%] & $50.5^{+5.1}_{-5.2}$ & $47.0^{+2.6}_{-5.1}$ \\
$f^R_{\text{N}}$ [\%] & $41.9^{+5.5}_{-5.5}$ & $45.4^{+5.5}_{-6.5}$ \\
$f^R_{\text{Si}}$ [\%] & $7.0^{+0.7}_{-0.9}$ & $7.0^{+0.7}_{-1.5}$ \\
$f^R_{\text{Fe}}$ [\%] & $0.39^{+0.26}_{-0.15}$ & $0.43^{+0.25}_{-0.16}$ \\
\hline
\textbf{Pure-proton population} & & \\
\hline
$\gamma^p$ & $-1.66^{+1.18}_{-1.34*}$ & $-0.05^{+0.66}_{-0.88}$ \\
$R^p_{\max}$ [EV] & $7.18^{+2.82}_{-1.96}$ & $10.01^{+2.95}_{-2.30}$ \\
$m^p$ & $6.1^{+0.5}_{-0.9}$ & $2.5^{+1.6}_{-3.7}$ \\
\hline
$f_p^{\text{obs}}(20~\mathrm{EeV})$ [\%] & $20.3^{+1.6}_{-1.4}$ & $19.4^{+1.4}_{-1.3}$ \\
$D/N_\text{dof}$ & 155.4/51 &  165.2/55 \\
\end{tabular}
\label{table 1}
\end{table}

Figure~\ref{fig: UHE_ga} shows the range of the predicted integral flux of UHE gamma rays observed on Earth from the KM3NeT-only and joint fits at the 68\% CL, along with upper limits from Auger \citep{auger_UHE_gamma_UL, auger_UHE_gamma_UL_2024} and the Telescope Array \citep{TA_UHE_gamma_UL}. No penalty was applied to the total deviance in either fit for this observable since all upper limits were far from being exceeded.
\par \smallskip

We obtain the following 68\% CL constraints on the source-evolution parameter of the pure-proton population $m^p$ and the observed proton fraction $f^{\text{obs}}_p(\text{20$\,\mathrm{EeV}$})$ from the KM3NeT-only fit:
\begin{align}\label{eq: km3net-only m}
    5.2 &\leq m^p \leq 6.6 \\[5pt]
    18.9\% \leq f_p^{\text{obs}} & (\text{20$\,\mathrm{EeV}$}) \leq 21.9\% \notag
\end{align}
and from the joint fit:
\begin{align}\label{eq: joint m}
    -1.2& \leq m^p \leq 4.1 \\[5pt]
    18.1\%\leq f_p^{\text{obs}} & ( \text{20$\,\mathrm{EeV}$}) \leq 20.8\% \notag
\end{align}
The remaining best-fit parameters are reported in Table~\ref{table 1}. 
\par \smallskip

In both fits, the observed proton fraction is tightly constrained by $\sigma(X_{\text{max}})$ at the upper bound and by $\langle X_{\text{max}}\rangle$ at the lower bound. Hence, our fits predict an observed proton fraction of $f_p^{\text{obs}}(\text{20$\,\mathrm{EeV}$}){\,\sim\,} 20\%$ under \textsc{epos-lhc}, which constrains the source-evolution parameter $m^p$ to the ranges above, under the two scenarios for the observed cosmogenic neutrino flux. In the scenario where only the KM3NeT dataset is considered, the detection of KM3-230213A favors a strong positive source evolution for the sources of UHE protons. In this scenario, $m^p$ is constrained at the upper bound by the IGRB flux, and at the lower bound by the non-zero event count in the KM3NeT dataset. In the scenario where the UHE neutrino datasets of Auger, IceCube, and KM3NeT are considered, $m^p$ is constrained to a wider range of lower values, including flat ($m^p=0$) and slightly negative source evolutions. In this scenario, $m^p$ is constrained at the upper bound by the zero event count in the Auger and IceCube datasets at the energy range of KM3-230213A, and at the lower bound by the non-zero event count in the KM3NeT dataset. 
\par \smallskip

Several high-luminosity (${\gtrsim} 10^{46}\,\mathrm{erg\,s^{-1}}$) classes of AGNs show a strong positive evolution of emissivity with redshift. AGNs tend to have a luminosity-dependent density evolution \citep{ueda2003cosmological, AGN_m_value}, whereby the growth of the source density with redshift and the redshift at which it peaks (corresponding roughly to $m$ and $z_0$, respectively, in Equation~\ref{eq: SE}) may depend on the luminosity class of the AGN. AGN classes with a source density evolution consistent with that of the pure-proton population in the KM3NeT-only fit (Inequality~\ref{eq: km3net-only m}) include: FSRQs with luminosity $10^{48}\,\mathrm{erg\,s^{-1}}$, whose density grows as $(1+z)^{7.4}$ up to $z{\,\sim\,}1.5$ and then declines as $(1+z)^{-6.5}$ \citep{FSRQ_m}; BL Lacs with luminosity $10^{47}\,\mathrm{erg\,s^{-1}}$, whose density grows as $(1+z)^{7.2}$ up to $z{\,\sim\,}1.3$ and then declines as $(1+z)^{-7.4}$ \citep{HL_BLlacs_m}; and high-luminosiy ($10^{46}{-}10^{47}\,\mathrm{erg\,s^{-1}}$) non-jetted AGNs, whose density grows as $(1+z)^{5.9}$ up to $z{\,\sim\,}1.9$ and then declines as $(1+z)^{-1.5}$ \citep{seyfert_m}. In addition to satisfying Inequality~\ref{eq: km3net-only m}, high-luminosity BL Lacs and FSRQs can produce a proton-dominated emission of UHECRs, if CRs are accelerated efficiently \citep{rodrigues2021active}, making them plausible candidates for a subdominant UHE proton population. Another candidate source of UHE protons is high-luminosity (${\gtrsim\,} 10^{50}\,\mathrm{erg\,s^{-1}}$) GRBs, whose jets are expected to be dominated by protons and neutrons in the classical fireball scenario \citep{horiuchi2012survival}. The cosmological evolution of GRBs has been linked to the SFR \citep[$m {\,\sim\,} 3$ in Equation~\ref{eq: SE};][]{GRB_SFR_m}, but other studies predict a weaker evolution of $m{\,\sim\,} 2$ \citep{GRB_m, GRB_m_2}, or a negative evolution of $m {\,\sim} -1$ \citep{GRB_m_neg}, all of which are consistent with the source evolution of the pure-proton population in the joint-fit (Inequality~\ref{eq: joint m}). Other source classes consistent with Inequality~\ref{eq: joint m} include: star-forming galaxies \citep[$m{\,\sim\,}3$;][]{SFR_m_1, SFR_m_value, SFR_2014}, low-radio-power AGNs \citep[$m{\,\sim\,}1$;][]{radio_gal_m}, low-luminosity ($10^{43}{-}10^{45}\,\mathrm{erg\,s^{-1}}$) non-jetted AGNs \citep[$m{\,\sim\,} 0{-}3$;][]{seyfert_m}, intermediate-luminosity ($10^{46}\,\mathrm{erg\,s^{-1}}$) BL Lacs and FSRQs \citep[$m{\,\sim\,} 0$;][]{FSRQ_m, HL_BLlacs_m}, and high-luminosity ($10^{46}{-}10^{47}\,\mathrm{erg\,s^{-1}}$) tidal disruption events \citep[$m{\,\sim\,} 0$;][]{TDE_flat_m}.
\par \smallskip

Neutrino and gamma-ray observations also constrain the spectral features of the pure-proton population. Harder proton spectra are favored since they produce fewer low-energy protons and hence fewer low-energy interactions. Here, we refer specifically to electron-positron pair production interactions with the CMB and photo-pion production interactions with the EBL. The former is a significant contributor to the diffuse GeV$-$TeV gamma-ray flux, which is constrained by the \textit{Fermi}-LAT data, and the latter contributes to the neutrino flux at $10^{15}{-}10^{17}\,\mathrm{eV}$, which is constrained by the IceCube high-energy event samples. The maximum rigidity of UHE protons, $R^p_{\text{max}}$, is constrained to values below the GZK limit of ${\sim\,} 50\,\mathrm{EeV}$ \citep{greisen1966end, zatsepin1966upper} in both fits by the measurements of $\sigma(X_{\text{max}})$, which disfavors a pure-proton composition at the highest energy bins, and the null observations of neutrinos beyond the energy range of KM3-230213A, which are expected to be abundant if UHE protons exceed the GZK limit \citep{kalashev2002ultrahigh}. These findings on the spectral index and maximum rigidity of a subdominant UHE proton component are in agreement with those of \cite{ehlert2024constraints}, who find that a hard proton spectrum with a sub-GZK maximum rigidity is most compatible with the UHECR and multi-messenger data. 
\par \smallskip

\section{Discussion}\label{sec: discussion}

The detection of a ${\sim\,} 0.2\,\mathrm{EeV}$ neutrino by the KM3NeT/ARCA detector could be the first observation of a cosmogenic neutrino. If interpreted as cosmogenic, the KM3-230213A event could constrain two important properties of UHECRs: the observed fraction of UHE protons and the cosmological evolution of their sources. We have adopted a scenario in which the UHECRs observed on Earth originate from two independent populations extending to $z_{\text{max}}=4$. The first population is mixed in composition, while the second population is comprised of UHE protons that contribute subdominantly to the CR spectrum. We used KM3-230213A and multi-messenger upper limits to place constraints on the UHE proton population. We found that KM3-230213A and multi-messenger upper limits only constrain the cosmological evolution of the sources of UHE protons, whereas the observed proton fraction is primarily constrained by the UHECR data. Under the hadronic interaction model \textsc{epos-lhc}, the $X_{\text{max}}$ data constrain the observed proton fraction to $f_p^{\text{obs}}(20~\mathrm{EeV}){\,\sim\,} 20\%$, which in turn constrains the strength of the source evolution of UHE protons to $m^p = 6.1^{+0.5}_{-0.9}$ when considering only the KM3NeT UHE neutrino dataset, and to $m^p = 2.5^{+1.6}_{-3.7}$ when the null observations of Auger and IceCube are taken into account. 
\par \smallskip

In this work, we have ignored source-to-source variations in a single population. It has been shown that UHECR sources act as standard candles since the source-to-source variation in the maximum rigidity is surprisingly low \citep{ehlert2023curious}. We have also assumed a homogeneous distribution of sources. This is a reasonable assumption given that the sources of UHECRs are unknown, and that a minimum local source density of ${\sim\,}10^{-5}\,\mathrm{Mpc^{-3}}$ is expected of UHECR sources \citep{abreu2013bounds}. Most importantly, the source distribution is expected to be homogeneous at very large distances, where the overwhelming majority of cosmogenic neutrinos and photons are produced. Hence, a local overdensity (or void) of sources has a negligible effect on the total flux of cosmogenic secondaries. For the case of UHECRs, local sources within ${\sim\,}80\,\mathrm{Mpc}$ account for nearly all of the observed flux beyond $10^{20}\,\mathrm{eV}$, but contribute significantly less at lower energies in the absence of extragalactic magnetic fields \citep{taylor2011need}. Such local contributions can induce deviations from the flux expected for a homogeneous source distribution \citep{ahlers2013ensemble}. In \citet{halim2023constraining}, the assumption of a homogeneous source distribution in fits to the spectrum and composition of UHECRs was tested by implementing different models of the local (${\lesssim\,}100\,\mathrm{Mpc}$) overdensity of matter. Two models were based on the SFR and stellar mass of local galaxies \citep{biteau2021stellar}, respectively, and another based on radio sources \citep{condon2002radio, condon2019radio}. They show no significant deviations from the homogeneous source distribution case. Similarly, source distributions following the local matter density reconstructed in \citet{hoffman2018quasi} also yield fits that do not significantly differ from the homogeneous source distribution case \citep{bister2024constraints}.
\par \smallskip

By working with one-dimensional simulations, we have implicitly assumed an isotropic emission of UHECRs by all sources. However, this is not the case for many jetted sources. The effect of jetted source emission on the observed spectrum was investigated in \citet{sippert2025effects}, who show clear deviations from the isotropic emission scenario when considering local sources and the presence of extragalactic magnetic fields. It could be argued that at sufficiently large distances, where the source distribution may be treated as homogeneous, jet orientations should also be uniformly distributed, so that the ensemble-averaged emission approaches the isotropic case. For the case of blazars (whose jets point directly towards Earth), they can be treated as a subset of the wider population of jetted AGNs \citep{urry1995unified}, particularly given their similar cosmological evolutions.
\par \smallskip

Another simplification implemented in this study is the omission of intergalactic magnetic fields. In the presence of magnetic fields, propagating particles experience more interactions due to longer trajectories. Moreover, UHECRs at the lower-energy end of the injection spectrum experience stronger magnetic deflections due to their lower rigidity, leading to diffusive propagation and limiting the distance from which they can reach the observer within a finite time, resulting in a suppression in their observed flux \citep[magnetic horizon effect;][]{aloisio2005anti}. Such effects are pronounced in strong (${>\,}1\,\mathrm{nG}$) magnetic fields and for low source densities (${<\,}10^{-4}\,\mathrm{Mpc^{-3}}$), as was investigated for the case of simple turbulent isotropic fields \citep[e.g.,][]{mollerach2020extragalactic, Gonzalez:2021ajv}, and more realistic configurations of the magnetic field following the cosmic web structure \citep[e.g.,][]{batista2014diffusion, wittkowski2017reconstructed}. In the extreme case of highly magnetized (${\sim\,}\mathrm{nG}$) voids in the cosmic web, uncertainties of up to 80\% and 5\% are predicted on the observed UHECR spectrum and composition, respectively, for the same injected spectrum \citep{alves2017implications}. The inclusion of intergalactic magnetic fields does not strongly modify the diffuse UHE and GeV$-$TeV gamma-ray flux \citep{ gelmini2008gzk, decerprit2011constraints}, but the cosmogenic neutrino flux could increase by a factor of a few at around ${1\,}\mathrm{EeV}$, for a mixed-composition injection \citep{wittkowski2019flux}. However, in scenarios where sources are uniformly distributed with separations smaller than the relevant propagation scales, the observed spectrum is expected to approach a universal shape, largely independent of magnetic field properties \citep[the propagation theorem;][]{aloisio2004diffusive}. In this analysis, we have implicitly assumed high source densities (or weak magnetic fields), and hence the validity of the propagation theorem. An investigation of alternative scenarios is beyond the scope of this paper.
\par \smallskip

Our results carry systematic uncertainties associated with our choice for the EBL model and photodisintegration cross sections, since they influence the interaction probability of UHECRs during propagation. Adopting an EBL model with a higher intensity \citep[e.g.,][]{dominguez2011extragalactic} compared to the \citet{gilmore2012semi} model used here could increase the cosmogenic neutrino flux by up to $25\%$ in the $10^{17}{-}10^{18}\,\mathrm{eV}$ range for hard, heavy-composition injection scenarios \citep{batista2019secondary}. Similarly, using PSB cross sections \citep{PSB_1976, PSB_1999} in place of TALYS can enhance the neutrino flux by up to $25\%$ below $10^{16.5}\,\mathrm{eV}$. The cosmogenic neutrino flux from protons, on the other hand, can be significantly reduced ($20{-}50\%$) below ${\sim\,}10^{16.5}\,\mathrm{eV}$ if the EBL model of \citet{dominguez2011extragalactic} is assumed \citep{batista2019secondary}. A similar reduction is found when using the EBL model of \citet{kneiske2004implications} compared to the less intense EBL of \citet{stecker2006intergalactic, stecker2007erratum}, whereas the GeV${-}$TeV gamma-ray flux between $10^{11}$ and $10^{13}\,\mathrm{eV}$ shows a slight increase \citep{decerprit2011constraints}.
These variations in the neutrino flux remain smaller than our model's statistical uncertainty. Nonetheless, adopting different EBL and photodisintegration models does have a non-trivial effect on the observed UHECR spectrum and composition \citep{batista2015effects}, which could influence the shape of the subdominant UHE proton component and the observed proton fraction. Investigating such effects is beyond the scope of this work.
\par \smallskip

The values for the spectral index and the maximum rigidity of any injected spectrum that best fit the measured UHECR spectrum depend on the shape of the energy cutoff. Functions that model the energy cutoff include a simple exponential, a broken exponential, which is equal to 1 if $E< ZR_{\text{max}}$ and $\exp \left(1 - E/ZR_{\text{max}}\right)$ if $E\geq ZR_{\text{max}}$, or a hyperbolic secant given by $\mathrm{sech}[(E/ZR_{\text{max}})^{\Delta}]$, where $\Delta$ relates to the steepness and width of the energy cutoff. In \citet{halim2023constraining} and \citet{bister2024constraints}, the measured UHECR spectrum is fitted using different energy cutoffs. They have shown that simple and broken exponential cutoffs require very hard injected spectra ($\propto E^{-\gamma}$ where $\gamma {\,\sim} -2$) to reproduce the observed energy spectrum, as is the case in this work. In contrast, sharper cutoffs, such as a hyperbolic secant with $\Delta=2$, fit the data with much softer spectra ($\gamma {\,\sim\,} 0.3{-}0.7$), and even softer ($\gamma {\,\sim\,} 1$) for $\Delta=3$, closer to expectations from known acceleration processes \citep[see also][]{Comisso:2024ymy}. Despite this, the authors have shown that the fit quality remains superior for simple and broken exponential cutoffs. Various studies have demonstrated that the spectral index is anti-correlated with the source-evolution parameter $m$ \citep{taylor2015indications, aab2017combined, heinze2019new, batista2019cosmogenic}. This implies that a softer injection spectrum of UHE protons, resulting from choosing a steeper energy cutoff, could lower the best-fit value of $m^p$ found in this work. 
\par \smallskip

Several studies have investigated the possibility that KM3-230213A originated from a steady cosmogenic neutrino component. It was demonstrated in \citet{km3_cosmogenic_companion} that a mixed-composition population of UHECRs propagating from a redshift of $z_{\text{max}}=6$, possibly with an additional subdominant proton component at UHEs, can contribute partially to the diffuse neutrino flux associated with the detection of KM3-230213A. Another implementation of the subdominant proton population scenario is presented in \citet{km3_alessandro}, where it is shown that a hard subdominant proton spectrum with a maximum rigidity of 100$\,\mathrm{EeV}$ can explain KM3-230213A without violating any multi-messenger limits if the source emissivity evolves as $(1+z)^3$ up to $z_{\text{max}}=5$. Their results partially align with our joint fit; however, whereas we investigate the entire parameter space, their analysis is limited to two scenarios for the proton spectrum, neither of which fully resembles our best-fit proton spectrum, making a direct comparison difficult.
\par \smallskip

Similar to this work, \citet{
muzio2025emergence} jointly fits the UHECR and neutrino data but under two scenarios for the true energy of the detected neutrino: $100\,\mathrm{PeV}$ and $1 \,\mathrm{EeV}$. They conclude that a $100 \,\mathrm{PeV}$ neutrino can be explained by source neutrinos, whereas a $1 \,\mathrm{EeV}$ neutrino strongly suggests a cosmogenic origin and hence the presence of an additional subdominant UHE proton population from which the observed neutrino was produced. Unlike this work, however, the authors assume a source evolution following the SFR up to a maximum comoving distance of 4.2 Gpc ($z{\,\sim\,}1.5$) throughout their study and do not report the quantitative constraints KM3-230213A places on their model parameters.
\par \smallskip

Another attempt to interpret the UHE neutrino event in light of UHECR observations was made using the Telescope Array (TA) data \citep{kuznetsov2026ultra}. The authors refer back to two fits for the UHECR spectrum and composition reported by the TA collaboration \citep{Bergman:2021ej} and show that these are compatible with the detection of KM3-230213A and the null observations by Auger, IceCube, and the Baikal-GVD experiment, at approximately the $2\sigma$ level. The TA fits predict a composition dominated by protons above the ankle. This naturally enhances the production of cosmogenic neutrinos at ${\sim\,} 1\,\mathrm{EeV}$, resulting in better agreement with the KM3NeT dataset. However, such a composition is not compatible with the $X_\mathrm{max}$ data reported by Auger.
\par \smallskip

\section{conclusion}\label{sec: conclusion}
We have used KM3-230213A and multi-messenger upper limits to place constraints on an UHE proton population that contributes subdominantly to the observed UHECR spectrum above the ankle. We find that the detection of a single neutrino event at the energy range of KM3-230213A solely with the KM3NeT exposure requires that the sources of the currently undiscovered UHE proton population must have a strong positive cosmological evolution, equivalent to the evolution of high-luminosity FSRQs, BL Lacs, and non-jetted AGNs.
On the other hand, including the null observations from the Pierre Auger and IceCube observatories disfavors sources with strong positive cosmological evolutions as the origin of the UHE proton population. Instead, sources with a moderately positive, flat, or slightly negative evolution become favorable, such as GRBs, star-forming galaxies, low-radio-power AGNs, low-luminosity non-jetted AGNs, and intermediate-luminosity FSRQs and BL Lacs. In both scenarios, the $X_{\text{max}}$ data constrain the observed proton fraction of UHECRs to be ${\sim\,} 20\%$ at 20$\,\mathrm{EeV}$.
\par \smallskip

Observations by future UHECR and neutrino observatories will significantly improve our results. One of the major goals of the Pierre Auger Observatory upgrade, the AugerPrime, is to assess the possible existence of a proton component at the highest energies \citep{augerprime_19}, and hence determine the fraction of protons in observed UHECRs. The presence of a proton component at the highest energies would improve the prospects for UHECR astronomy, since protons are less strongly deflected by magnetic fields than heavier nuclei with the same energy. At the same time, the ongoing operation and construction of current neutrino telescopes such as KM3NeT/ARCA \citep{km3net2024astronomy} and Baikal-GVD \citep{Baikal-GVD_2023}, and the planned operation of next-generation neutrino telescopes, such as GRAND \citep{GRAND_2020}, RNO-G \citep{RNO-G_2021}, IceCube-Gen2 \citep{icecube_gen2}, P-ONE \citep{P-one_2024}, NEON \citep{NEON_2025}, TRIDENT \citep{TRIDENT_2023}, and HUNT \citep{HUNT_2026}, will substantially increase event statistics, allowing for a more precise determination of the cosmogenic neutrino flux.

\begin{acknowledgments}
AA, AvV, and ST acknowledge support from Khalifa University’s internal grants, RIG-S-2023-070 and RIG-2024-047. The authors gratefully acknowledge the KM3NeT collaboration for many valuable discussions and constructive feedback on the results. The authors would like to thank the anonymous referee for useful comments and careful reading of the manuscript. AA would like to thank Ralph Engel, Maximilian Meier, and Sergey Troitsky for useful discussions. AA would also like to thank Sergio Petrera for providing parameter values of the parameterization in \citet{abreu2013interpretation} for the \textsc{epos-lhc} model.
\end{acknowledgments}

\begin{contribution}
AA led the analysis and was responsible for writing and submitting the manuscript. AvV developed the idea for this study. DE assisted with the analysis and idea development. ST provided data and assisted with the idea development. All authors equally contributed to reviewing and editing the manuscript.
\end{contribution}

\footnotesize
\bibliography{sample701_arxiv}
\bibliographystyle{aasjournalv7}

\end{document}